\magnification=\magstep 1     
\noindent     
{\bf Entropy Defined, Entropy Increase and Decoherence Understood, and
Some Black-Hole Puzzles Solved}  
\bigskip     
\noindent  
Bernard S. Kay, Department of Mathematics, University of York, 
York YO1 5DD, UK     
\bigskip    
\bigskip    
\noindent  
{\bf Statistical mechanics explains thermodynamics in terms of (quantum)
mechanics by equating the entropy of a microstate of a closed system 
with the logarithm of the number of microstates in the macrostate to
which it belongs, but the question `what is a macrostate?' has
never been answered except in a vague, subjective, way. However
Hawking's discovery$^{1,2}$ of black hole evaporation led to a formula
for black hole entropy with no subjective element. In this letter, we
argue from this result, together with the assumption$^3$ that `black
hole thermodynamics is just ordinary thermodynamics applied to black
holes', that a macrostate for a general (quantum gravitational) closed
system is {\it an equivalence class of matter-gravity  microstates with
the same expectation values for the matter degrees of freedom alone}. 
Not only does this finally answer the question `what is entropy?',
but it also predicts the equality of the thermodynamic entropy of a
black hole  with the matter and the gravity entropy-like quantities
derived$^{4,5}$ from the Euclidean path integral.  Furthermore it gives us a
clear glimpse of an ultimate synthesis of quantum theory and gravity in
which we see that (a) gravity acts as a {\it universal environment},
thus predicting that, if the initial state of the universe is
unentangled, its entropy must go on increasing forever, (b) {\it the
gravitational field has degrees  of freedom, but no observables}, thus
enabling gravity to perform the trick of providing {\it an objective
continual process of decoherence}. All the above rests on the validity
of unitarity. The `information-loss puzzle$^6$' had raised doubts about
that. But we suggest a resolution for this puzzle.} 

Our thesis is that thermodynamics can be derived as a consequence of the
microscopic laws of nature only when these microscopic laws are taken to
be the laws of quantum gravity.  Relying only on basic, conservative,
assumptions about the quantum mechanical framework underlying quantum
gravity, together with well-established results from the subject of
(quantum) `black-hole thermodynamics',  not only do we then find that
entropy acquires an objective definition, but also a new paradigm for
the laws of nature emerges which seems capable of solving many of the
longstanding puzzles of theoretical physics and cosmology. Moreover, the
new paradigm corrects our previous understanding of black hole
thermodynamics in a way which explains some previously unexplained
coincidences.

We must first clarify what we mean by a closed system in quantum
gravity.  Because of  interactions, and because gravity is `universal',
i.e. it couples to all matter, this can only be a model which either is
intended  literally to be a description of the whole universe, or which
describes our system of interest {\it as if} it were the whole universe.
For example, if one were interested in the physics of a single neutron
star, one would consider a fictitious universe consisting of a single
neutron star sitting in an empty, apart from quantum vacuum
fluctuations, asymptotically flat space. Fictitious systems in `boxes'
are also permitted, with due regard to appropriate fictitious boundary
conditions, in the same spirit.

Our starting point is to posit the existence, and correctness, of a
quantum gravity theory which is in conformity with standard quantum
mechanics in the sense that the following two assumptions hold:  First,
the theory assigns independent dynamical degrees of freedom for gravity
and matter so the total Hilbert space for a closed system will take the, 
tensor product, form
$$H=H_{\hbox{matter}}\otimes H_{\hbox{gravity}}.$$  
Second, any full description of any closed system corresponds to the
standard quantum mechanical notion of a {\it single microstate} -- i.e. to a
density operator $\rho$ which takes the form  $|\Psi\rangle\langle\Psi|$
where $|\Psi\rangle$ is a vector, or `total wave function' in $H$; in
other words, to what is usually called a `pure total state'.
We shall postpone the question of `what is time' to after our discussion of
entropy increase below, and, as an interim interpretation, simply
think of the total Hilbert space as representing the possible microstates
of the system `at a given single time'.

By standard quantum mechanics$^7$, we know that, given any total Hilbert
space $H$ which takes the form $H_a\otimes H_b$, (we are interested in
the case where $a$ stands for {\it matter} and $b$ for {\it gravity})
and a microstate $\rho=|\Psi\rangle \langle\Psi|$, then, if we are only
concerned about an observable $A$ tied to the $a$ degrees of freedom,
the expectation value $\langle A\rangle$ of $A$ is given by a density
operator $\rho_a$ on $H_a$, namely the `partial trace' of $\rho$ over
the Hilbert space $H_b$, satisfying
$$\hbox{trace}(\rho A\otimes I_b)_H=
\hbox{trace}(\rho_a A)_{H_a}\eqno{(1)}$$
where $I_b$ stands for the identity operator on $H_b$. Similarly there
will be $\rho_b$ on $H_b$, suitable for observables $B$ tied to the $b$
degrees of freedom, and  by the standard von Neumann procedure$^7$, we
can assign to each of these partial states the `partial entropies'
$S_a=-\hbox{trace}(\rho_a\log\rho_a)$ and 
$S_b=-\hbox{trace}(\rho_b\log\rho_b)$, which, as we explain below, and
as is well known, must (as a consequence of the purity of the total
state) necessarily be equal (and, of course, non-negative).  Of course,
the total von Neumann entropy of the microstate $\rho$, will be zero,
and when we consider a time-evolution, below, will  remain zero for all
time.  Thus  `additivity of the entropy' fails to the maximum possible
extent here. {\it But in the sequel, the von Neumann entropy of the
total state will have no direct physical interpretation}.  $\rho_a$ is
the proper mathematical representation of what we call here an `$a$ {\it
macrostate}'.  In the case $\rho_a$ is a projector onto an $n$-dimensional
subspace, then one may identify the {\it `number of microstates in the
macrostate $\rho_a$'} with the number $n$.  But, we shall continue to
use this phrase, picturesquely, also to describe the general case.  This
way we can think of the $a$ macrostates as {\it the equivalence classes
of microstates which are indistinguishable from one another,  as far as
the $a$ observables are concerned}.  The partial entropy $S_a$ may 
then be thought of as the logarithm of the number of microstates for the
total system contained in the macrostate $\rho_a$. A corresponding
statement holds for the interpretation of $S_b$ ($=S_a$).

It will be helpful to recall how the equality $S_a=S_b$ may be proven:
The total wave function $|\Psi\rangle$ may be written as a superposition, with
coefficients $c_i$ satisfying  $\sum_i |c_i|^2=1$, of single tensor
products  
$$|\Psi\rangle=\sum_i c_i|\alpha_i\rangle\otimes|\beta_i\rangle$$  
where the $|\alpha_i\rangle$ and $|\beta_i\rangle$ (which depend on 
$|\Psi\rangle$) are orthonormal sets of basis vectors in $H_a$ and $H_b$
respectively.  One easily has from this that $$\rho_a=\sum_i
|c_i|^2|\alpha_i\rangle\langle\alpha_i|\quad\hbox{and}\quad
\rho_b=\sum_i |c_i|^2 |\beta_i\rangle\langle\beta_i|$$  and from this it
easily follows that 
$$S_a=S_b=-\sum_i|c_i|^2\log|c_i|^2.$$ 
Finally, we remark that when the total wave function can be written as a
single non-zero tensor product, the total state is often said to be
unentangled.  In this case, the partial entropies will be zero.  For
this reason, the value of the (equal) partial entropies in the general
case, is often referred to as `entanglement entropy'.

Returning to the case of interest, where $a$ stands for {\it matter} and
$b$ for {\it gravity}, we shall now argue that the value of the
{\it physical entropy}, $S$, of a matter-gravity microstate
$\rho=|\Psi\rangle \langle\Psi|$ must be the value of the (equal) partial
matter and gravity entropies of $\rho$, i.e. it must be its `matter-gravity
entanglement entropy'. In particular,
$$S=S_{\hbox{matter}}\quad (=
-\hbox{trace}(\rho_{\hbox{matter}}\log\rho_{\hbox{matter}}))\eqno{(2)}$$
or, in words:
\smallskip
\noindent
{\it The physical entropy, $S$, of the microstate $\rho$ is the
logarithm of the number of matter-gravity microstates for which the
expectation values of all the matter degrees of freedom coincide with
the expectation values in $\rho$}.
\smallskip
More precisely we shall argue that, if there is any quantity
at all in quantum gravity, which is a function on the set of all
possible microstates, taking values in the non-negative real numbers,
and which coincides with what we call `entropy' both in every-day
and in black-hole thermodynamics, then it must be given by (2).

We shall give two entirely independent arguments for this
statement. We feel that, while each argument, by itself, might be open
to some reasonable doubt, the two arguments taken together make a very
strong case that (2) is true. Our first argument begins by observing
that, in the thermodynamics of everyday closed systems, one is concerned
with measurements of the matter, but not of the gravity, degrees of
freedom. Thus the (vague, ill-defined) old statistical-mechanical
entropy is equal to
\smallskip 
\item{(I)} 
the entropy one would obtain, by modifying the von Neumann formula with
the usual `coarse graining procedures', when one coarse-grains in a way
appropriate to measurements of the matter degrees of freedom with a
given degree of experimental accuracy, and gravity is not measured at
all. 
\smallskip  
\noindent
On the other hand, (2) asserts that the physical entropy is 
\smallskip 
\item{(II)} 
the entropy obtained when one `coarse grains' as if measurements of the
matter degrees of freedom could be made (to within the fundamental
limitations of the Heisenberg uncertainty principle) with {\it maximal}
accuracy, and gravity is {\it still} not measured at all. 
\smallskip 
\noindent
Because of the well-known robustness of the traditional arguments of
statistical mechanics under changes in the notion of coarse graining, 
it seems plausible that, as long as there's a reasonable number of
gravitational degrees of freedom, it will not make any difference which
of these two coarse grainings we actually use. If it did, thermodynamics
would start going wrong when one started to make measurements of matter
too accurately.  Thus, we conclude that in every-day thermodynamics,
while we have a lot of choice as to how entropy may be defined, the
entropy {\it may be taken to be} the $S$ of (2), and this is the only
definition which is free from any subjective element; {\it gravity
provides an ideal, and objective, notion of coarse-graining}. 
But in the case of a {\it black hole} we combine our knowledge of
Hawking's formula {\it temperature equals surface gravity over $2\pi$}
with the hypothesis$^3$ that `black hole thermodynamics is ordinary
thermodynamics applied to black holes' to conclude that the entropy
must equal
\smallskip
\item{(i)} the entropy obtained using the thermodynamic formula
$dE=TdS+\hbox{\it work terms}$, from the above temperature formula.
\smallskip
\noindent
This takes the value$^2$ `one quarter of the area of the event horizon'
and the important point for our argument is that this is  a definite
value, free from any subjective element. We conclude that if we are to
have a single formula valid for {\it all} closed systems, both every-day
and black-hole, the physical entropy has to be given by (2).

Our second argument takes as its starting point the remarkable
result$^4$ from the `Euclidean path-integral approach$^5$' to quantum
black holes that another `entropy-like quantity' namely
\smallskip
\item{(ii)} the `entropy' obtained$^{4,5}$ from 
the pure gravity partition function$^{4,5}$ for a quantum black hole
\smallskip
\noindent
turns out, by a hitherto unexplained coincidence, to equal, to
zero-loops$^{4,5}$, one quarter of the area of the event horizon, i.e.
the same value as (i). On the other hand, we shall argue below that (ii)
ought (to zero loops) to be equated with $S_{\hbox{gravity}}$ which we
know must equal $S_{\hbox{matter}}$. Thus we conclude that (2) is true
for the case of a black hole.  By considering examples of closed systems
which include both black holes and every-day objects, one can further
argue that this conclusion must extend to general microstates. 

Once one accepts the definition (2) for physical entropy, the prospect
opens up of being able, once we have a theory of quantum gravity,  to
replace the traditional, partially unsatisfactory, explanation of
thermodynamics in terms of a (quantum) `statistical mechanics',  by an
explanation in terms of a quantum-gravitational `mechanics'. In fact, we
can already see the qualitative shape such an explanation will take: The
laws of thermodynamics should emerge for the {\it matter degrees of
freedom} regarded as an {\it open system}  in contact with the {\it
environment} provided by the gravitational field, and coupled to the
latter according to the interactions determined by the physically
correct full theory of quantum gravity when the entropy is regarded as
the attribute of each microstate of the total {\it matter-gravity} {\it
closed system} defined by (2). The fact that the gravitational
coupling is so weak, encourages us to expect that an explanation along
these lines will be successful.

We have now arrived at the right place to discuss the, interrelated,
problems of `time' and `entropy increase'. In order to proceed with the
enterprise outlined above along anything like traditional lines, one
will need a notion for a flow of time, represented by a one-parameter
family of  unitary operators, which we shall call a {\it flow}
$t\mapsto U(t)$ on $H$ with $t$ ranging over the non-negative real
numbers, mapping any `initial microstate'
$\rho_0=|\Psi_0\rangle\langle\Psi_0|$ at some `initial time' $t=0$ into
the microstate $\rho_t=|\Psi_t\rangle\langle\Psi_t|$ at a `later time',
$t$, according to the transformation 
$$\rho_t=U(t)\rho_0 U(t)^{-1}.\eqno{(3)}$$ 
So a single microstate at one time evolves to a single  microstate at a
later time. It is very important to realize that it will be the theory
of quantum gravity which will, presumably, tell us which  such flows
deserve to be called {\it time-flows}. (In fact, it seems conceivable that
a quantum theory of gravity {\it is} just a specification of which flows
are time-flows.) The problem of the increase of entropy of a closed
system seems now susceptible of solution, once we add the assumption,
for any closed system of interest, that the initial matter-gravity
microstate is totally unentangled.  Its initial entropy will then
obviously be zero, and it is obvious, that, at least at the start, its
entropy $S(t)$ can only increase.  But we should be able to say much
more: There would seem to be very good hopes that arguments similar to
the traditional arguments of statistical mechanics for open systems in
contact with environments, but now of a deterministic kind, will, for
the time-flows sanctioned by quantum gravity, now ensure that $S(t)$
must go on increasing for ever and can never turn around and start
decreasing at some time.  In particular, given such a notion of
time-flow, if we assume the initial microstate $\rho_0$, has zero, or
low, entropy then one should be able to argue that {\it the entropy of
the universe must go on increasing forever}.  This way of explaining
entropy increase is to be contrasted with attempts within the
traditional paradigm.  While something like the above picture can be
argued for with a suitable notion of coarse-graining, as long as one
believed that, ultimately, physical entropy should be identified with
the von Neumann entropy of the total state, then, assuming the universe
to be a single microstate and its time evolution to be unitary, such
explanations were open to the well-known objection that, in a fully
accurate description, the  entropy must remain zero for all time.

Note that, conceptually, $t$ above should presumably be thought of just as a
`parameter'.  Rather, it is tempting to identify the {\it time} at which a
given unitary $U$ ($=U(t)$ for some $t$, for some time-flow) {\it occurs},
with the entropy of $U\rho_0U^{-1}$.  This remark seems to offer the
propect of at least a partial solution to the `problem of time in
quantum gravity'.

Next, in order to remove an apparent obstacle to the validity of our
thesis, we need to discuss in detail the nature of the quantum state of
matter-gravity appropriate for the description of a black hole, as we
now explain.  Above, {\it we} have been taking it for granted that a
black hole will always be correctly described by a single microstate of
matter-gravity. {\it In the traditional picture} it is taken for
granted, instead, that a black hole {\it cannot} be described by a
single microstate.  It is claimed, on the contrary, that it resembles a
macrostate in that it needs to be described by a total density operator
$\rho$ with (total) non-zero (von Neumann) entropy, corresponding to a
statistical mixture (i.e. a sum with suitable coefficients) of single 
microstates of the form $\rho_i= |\Psi_i\rangle\langle\Psi_i|$.  As
mentioned at the outset, this contradicts unitarity (in fact it's not
really clear what its physical interpretation could be) and this
contradiction is known as the `information loss puzzle$^6$'. To clarify
this, we need to distinguish between:
\smallskip
\item{(a)} the case of a black hole which is freshly formed from a
collapsing mini-star modelled by a state with zero matter-gravity
entanglement in the distant past, and just beginning to evaporate, and
\smallskip
\item{(b)} the stable equilibrium situation made possible by keeping the
black hole in a suitably small box.
\smallskip
\noindent
It seems that one of the main (at least implicit) reasons for the belief
in the `information loss puzzle' is that it was also taken for granted
that, in case (b) -- the perfect equilibrium situation the radiating
black hole is, at the start,  `striving' to attain -- the only
physically possible equilibrium (total) states for this system are
completely thermal.  Instead, we shall now argue that the possible
equilibrium (total) states of this system should equally include
matter-gravity  microstates in which the matter degrees of freedom and
the gravitational degrees of freedom are separately in thermal
equilibrium, but entangled. In other words that they should include
single microstates which resemble  `{\it double-KMS states}$^{8,9}$' as
we explain below.  We suggest that this argument deserves to be
considered as a resolution of the `information loss puzzle' in that it
removes the main reason for believing there to be such a puzzle.

To see that one expects there to be such single equilibrium microstates,
we consider a simple analogy.  Instead of matter and gravity, imagine a
couplet of two real free scalar fields -- called say the a-on field and
the b-on field -- in Minkowski space. Then, one knows that the
generating functional (for all the n-point Wightman functions) of a
thermal total state, at inverse temperature $\beta$, is given explicitly
by: 
$$\langle \exp(ia(f)+ib(g))\rangle=\exp\lbrace-{1\over 2}\langle
Kf|\coth(\beta h/2)Kf\rangle -{1\over 2}\langle Kg|\coth(\beta
h/2)Kg\rangle\rbrace$$ 
where $a(f)$ and $b(g)$ are integrals over Minkowski space of the fields
with test functions $f$ and $g$, $Kf$ and $Kg$ denote the usual
restrictions of the Fourier transforms of $f$ and $g$ to the future 
mass shell, $h$ is the usual free field `one-particle Hamiltonian',
and the inner products in the exponent are defined by the
usual Lorentz-invariant integral.  However it is clear, from the work in
refs. 8 and 9, that, aside from this total thermal (`KMS') state, one
can also have the `double-KMS' state 
$$\langle \exp(ia(f)+ib(g))\rangle=\exp\lbrace-{1\over 2}\langle
Kf|\coth(\beta h/2)Kf\rangle+ \hbox{Re}\langle Kf|\hbox{cosech}(\beta
h/2)Kg \rangle$$ $$-{1\over 2}\langle Kg|\coth(\beta
h/2)Kg\rangle\rbrace$$ 
which has non-trivial correlations, i.e. entanglement, between the a-on
field and the b-on field which are such that the total state is pure,
i.e. a single microstate, but of course the partial states for the a-on
field and the b-on field are identical to those of the thermal state and
remain thermal.  Note that this double-KMS state will have Green
functions which are `periodic in imaginary time', a property which is
sometimes, mistakenly, believed to be exclusive to thermal states. Note
also that one can give this model an amusing interpretation where the
a-on field and the b-on field are identified with the restrictions of a
single real scalar field to the left wedge and the right wedge of a
fictitious Minkowski space, and the total (`double KMS') state is then
analogous to the Minkowski vacuum state. However, except in the
two-dimensional massless case, where this identification can be achieved
exactly by conformally mapping each of the wedges (in such a way that
the wedge-preserving Lorentz boosts go over, in each case, into the
forwards Minkowski time evolution), if one tried to take this
identification literally, one would get$^8$ a  strange `one-particle
Hamiltonian'.

Pursuing our analogy, {\it a-ons for matter, b-ons for gravity}, our
suggestion, then, is that, also for a case (b) black hole, there will be
a single microstate which resembles these double KMS states and this
will be the physically relevant equilibrium state; {\it not} a totally
thermal state.  Concomitantly, we suggest that the full matter-gravity
Euclidean path integral$^5$  with the usual boundary conditions at the
box and the usual periodicity in imaginary time is the generating
functional for this single microstate, and that the traditional
interpretation as a total matter-gravity thermal partition function is
incorrect. Note however that, also with this new interpretation, if one
takes the partial trace by integrating out all the matter fields,
one will clearly still obtain a thermal  partition function for the
gravity degrees of freedom alone which coincides with that of refs. 4
and 5 to `zero-loops'.  Hence we arrive at the conclusion that
$S_{\hbox{gravity}}$ is, to zero-loops, equal to (ii), as we assumed in 
our second argument for the validity of (2). We remark (A) If one were
content to let our first argument for the validity of (2) stand alone,
then one could turn that second argument into an {\it explanation} as to
why  (i) and (ii) approximately coincide. Moreover,  since
$S_{\hbox{matter}}=S_{\hbox{gravity}}$, we arrive at the {\it
prediction} that the third entropy-like quantity
\smallskip
\item{(iii)} the `entropy' derived from the matter partition function
obtained from the full matter-gravity Euclidean path integral by
integrating out the gravity field
\smallskip 
\noindent
must also coincide, to a good approximation, with (i) and (ii).  Prior to the
present work, it was not clear that, or why, (i), (ii) and (iii) should have
similar values and it was not clear which of (i), (ii), (iii), or
possibly the sum of some of these, is the physical entropy of a black
hole.

As further support for this suggestion, note that it would anyway seem
unreasonable for the equilibrium state to which a radiating black hole
is striving, to be a totally thermal state since, in such a state,
matter and gravity would be unentangled and, as has often been
explicitly or implicitly assumed, the matter and gravity partial von
Neumann entropies would sum to the total von Neumann entropy.  But,
surely, because of interactions, the collapse process will tend to
entangle an initially unentangled state leading to the failure of the
von Neumann entropies to be additive. So just on these grounds,  pure
double-KMS-like states would seem better candidates for the physically
relevant equilibrium states. 

Finally,  the assumptions we have made in this letter seem to tell us
both {\it that} and {\it how} gravity must play a role in the mystery of
the `collapse of the wave packet':  Our definition of entropy (2) may be
interpreted as indicating that {\it the degrees of freedom of the
gravitational field are unobservable.}   We now propose to elevate this
statement to a {\it principle}.   With this principle, we may identify
the algebra of observables with the algebra of operators $A$ on the
Hilbert space $H_{\hbox{matter}}$ and, by (1), the expectation value of
such an operator in the total microstate $\rho=|\Psi\rangle\langle\Psi|$
will be determined by the partial matter density operator
$\rho_{\hbox{matter}}$ through    
$$\langle A\rangle=\hbox{trace}(\rho_{\hbox{matter}}A).$$  
The physical entropy of $\rho$ then becomes the von Neumann entropy of
$\rho_{\hbox{matter}}$ and the time evolution (3) between two times will
act so as to transform a zero-entropy such partial density operator --
corresponding to a total initial microstate  which is unentangled --
into a density operator with non-zero, and by our arguments above, ever
increasing, entropy.  Thus, with our principle, as far as the (matter)
observables are concerned, the gravitational field must play a role
similar to that which would be played, in the traditional view, by a
combination of a unitary time-evolution and a continual process of
`measurement', i.e. the `reduction' part of measurement; perhaps the
`collapse' remains  mysterious. Previously there had been attempts to
understand such `decoherence' in  terms of `the environment' but they
did not really work because, as long as the environment consisted of
degrees of freedom which were observable in principle, the wave function
could, at best, be reduced `for all practical purposes', but never {\it
actually} be reduced.  Instead, by constituting an environment which is
in principle unobservable, the gravitational field gives us an {\it
absolute notion of reduction} thus finally offering at least a partial
explanation for why we don't see `Schr\"odinger cats' etc.  In the
`measurements' which are taking place in the decoherence  process, the
`cut$^7$' between the unitary time evolution and the `reduction'  is now
understood not to occur, as von Neumann argued, at some arbitrarily
located stage, between `system' and `observer', but rather to be
located, precisely, where the matter `open system' makes contact with
the `gravity environment'.  All this clarifies the foundations of
quantum mechanics in much the same way as our objective notion of coarse
graining clarified the foundations of statistical mechanics.  In fact,
with the above principle, we have achieved not only a clarification, but
also a {\it unification} of the two subjects; {\it entropy increase and
decoherence are seen to be two aspects of the same thing}.

Hawking's calculation$^1$ of black hole evaporation, twenty four years
ago, provided a crucial clue which has enabled us, here, to achieve
a new understanding of thermodynamics and to find a single explanation
for the origin both of entropy increase and of decoherence. 
Contradicting Einstein, Hawking$^6$ said that `God not only plays dice,
but sometimes throws the dice where we cannot see them'. Here, we have
confirmed that this is true, but have argued that the dice we cannot see
are nothing other than the degrees of freedom of the gravitational field
itself; while remaining hidden, they ensure the overall unitarity of
the theory.

The framework we have arrived at is remarkably close to conventional
quantum mechanics.  But it has some important new features: 
unobservable degrees of freedom, and a built-in irreversibility; time
has a beginning, and the initial state is unentangled.  The challenge it
poses is to construct a full quantum theory of gravity within this new
paradigm.

\bigskip

\item{1.} Hawking, S.W. Black hole explosions? Nature {\bf 248}, 30-31
(1974).

\item{2.} Hawking, S.W. Particle creation by black holes.
 Commun. Math. Phys. {\bf 43}, 199-220, (1975).

\item{3.} Wald, R.M. Black holes and thermodynamics. 
in {\it Proceedings of the Chandra Symposium: ``Black Holes and Relativistic
Stars''} (ed. Wald, R.M.) (U. of Chicago Press, to appear) (gr-qc/9702022)

\item{4.} Gibbons, G.W. \& Hawking, S.W. Action integrals and partition
functions in quantum gravity.  Phys. Rev. {\bf D15}, 2752-2756 (1977).
  
\item{5.} Hawking, S.W. The path integral approach to quantum gravity.
in {\it General Relativity, An Einstein Centenary Survey} (eds Hawking, S.W.,
Israel, W.) 746-785 (Cambridge University Press, 1979).

\item{6.} Hawking, S.W. Breakdown of predictability in gravitational
collapse.  Phys. Rev. {\bf D14}, 2460-2473 (1976). 
  
\item{7.} von Neumann, J. {\it The Mathematical Foundations of Quantum
Mechanics} (Princeton University Press, 1955).

\item{8.} Kay, B.S. The double wedge algebra for quantum fields on
Schwarzschild and\hfil\break 
Minkowski spacetimes.  Commun. Math. Phys. {\bf 100}, 57-81, (1985).

\item{9.} Kay, B.S. Purification of KMS states. Helvetica Physica Acta
{\bf 58}, 1030-1040 (1985).

\bigskip 
\noindent 
{\bf Acknowledgements.} 
I am very grateful to Leslie Sage for reading, and constructively
criticizing, a first version of this work and to Wanda Andreoni and
Alessandro Curioni who (during the week 15-22 January 1998) did the same
for successive further versions.  Especially, I am indebted to
Alessandro Curioni for reminding me that entropy is time's arrow and for
insisting that, of the two concepts, entropy is the more fundamental.
\bigskip
\noindent
Correspondence should be addressed to the author (e-mail bsk2@york.ac.uk).

\bye